\documentstyle[12pt]{article}

\def\beqa{\begin{eqnarray}}
\def\eeqa{\end{eqnarray}}
\def\beq{\begin{equation}}
\def\eeq{\end{equation}}

\def\vol{\int d^4x\,\sqrt{-g}}

\def\half{\frac{1}{2}}
\def\gu{g^{\mu\nu}}
\def\gd{g_{\mu\nu}}

\def\umunu{^{\mu\nu}}
\def\dmunu{_{\mu\nu}}

\def\da{_{\alpha}}

\def\uab{^{\alpha\beta}}

\def\ddemunu{_{;\mu\nu}}

\def\ddemu{_{;\mu}}  
\def\ddenu{_{;\nu}}  
\def\ddea{_{;\alpha}}  
\def\ddeb{_{;\beta}}

\let\lam=\lambda  

\let\gam=\gamma
\let\alp=\alpha
\let\sig=\sigma

\let\lb=\label
\renewcommand{\epsilon}{\varepsilon}
\let\no=\nonumber

\def\pr{{\it Phys. Rev.}\ }
\def\prl{{\it Phys. Rev. Lett.}\ }
\def\pl{{\it Phys. Lett.}\ }
\def\np{{\it Nucl. Phys.}\ }

\def\cqg{{\it Class. Quantum Grav.}\ }

\def\grg{{\it Gen. Relativ. Grav.}\ }

\def\apj{{\it Ap. J.}\ }
\def\aa{{\it Astron. Astrophys.}\ }

\def\araa{{\it Ann. Rev. Astr. Ap.}\ }

\def\f{F(\phi)}
\def\fp{F'(\phi)}

\def\p{\phi}
\def\pv{\varphi}
\def\v{V(\phi)}
\def\vp{V'(\phi)}
\def\l{\cal L}

\begin{document}
\renewcommand{\theequation}{\thesection.\arabic{equation}}
\begin{titlepage}
        \title{Newtonian limit of Induced Gravity}
\author{S. Calchi Novati\thanks{E-mail: novati@sa.infn.it},~~
S. Capozziello\thanks{E-mail: capozziello@sa.infn.it},~~
and G. Lambiase\thanks{E-mail: lambiase@sa.infn.it} \\
{\em Dipartimento di Scienze Fisiche "E.R. Caianiello"} \\
 {\em Universit\'a di Salerno, 84081 Baronissi (Sa), Italy.} \\
 {\em Istituto Nazionale di Fisica Nucleare, Sez. di Napoli, Italy.} \\ }
\date{\today}
\maketitle

\begin{abstract}
We discuss the weak-field limit of induced gravity and show that
results directly depend on the coupling and self--interaction potential of the
scalar field. A static spherically symmetric exact solution is found and its
conformal properties are studied. As  an application, it is shown that the
light deflection angle and the microlensing
quantities can be parametrized by the coupling of the theory.
\end{abstract}
\vspace{10.mm}
PACS number(s): 04.50.+h \\
\vspace{5.mm}
Keyword(s): scalar-tensor theories of  gravity,  Newtonian limit,
gravitational lensing.
\vfill
\end{titlepage}

\section{\bf Introduction}
\setcounter{equation}{0}

Extended theories of gravity have become a sort of paradigm in the
study of gravitational interaction since several motivations push
for enlarging the traditional scheme of Einstein general
relativity. Such issues come, essentially, from cosmology and
quantum field theory.

In the first case, it is well known that higher--derivative
theories \cite{starobinsky} and scalar--tensor theories \cite{la}
furnish inflationary cosmological solutions capable, in principle,
of solving the shortcomings of standard cosmological model
\cite{weinberg}. Besides, they have relevant features also from
the quantum cosmology point of view since they give interesting
solutions to the initial condition problem, at least in the
restricted context of minisuperspaces \cite{vilenkin}.

In the second case, every unification scheme as superstrings, supergravity or
great unified theories, takes into account effective actions where
nonminimal coupling to the geometry or higher order terms in the curvature invariants
come out. Such contributions are due to one--loop or higher--loop
corrections in the high curvature regimes near the full
(not yet available) quantum gravity regime \cite{odintsov}.
However, in the weak--limit approximation, all these classes
of theories should be expected to reproduce the Einstein general
relativity which, in any case, is experimentally tested only in this
limit \cite{will}.

This fact is matter of debate since several relativistic theories
{\it do not} reproduce Einstein results at the Newtonian approximation
but, in some sense, generalize them. In fact, as  it was firstly
noticed by Stelle \cite{stelle}, a $R^2$--theory gives rise to
Yukawa--like corrections to the Newtonian potential which could have
interesting physical consequences.

For example, some authors claim for explaining the flat rotation curves
of galaxies by using such terms \cite{sanders}. Others \cite{mannheim}
have shown that a conformal theory of gravity is nothing else but
a fourth--order theory containing such terms in the Newtonian limit and,
by invoking these results, it could be possible to explain the missing
matter problem without dark matter.

Besides, indications of an apparent, anomalous, long--range acceleration
revealed from the data analysis of Pioneer 10/11, Galileo, and Ulysses
spacecrafts could be framed in a general theoretical scheme by taking
into account Yukawa--like or higher order corrections to the Newtonian
potential \cite{anderson}.

In general, any relativistic theory of gravitation can yield corrections
to the Newton potential (see for example \cite{schmidt}) which in the
post-Newtonian (PPN) formalism, could furnish tests for the same theory
\cite{will}.

Futhermore the newborn {\it gravitational lensing astronomy} \cite{ehlers} is giving
rise to additional tests for general relativity over small, large, and very large scales
which very soon will provide direct measurements for the variation of Newton coupling
$G_{N}$ \cite{krauss}, the potential of galaxies and clusters of galaxies \cite{nottale}
and several other features of gravitating systems. Such data will be very likely capable
of confirming or ruling out the physical consistency of extended theories of gravity.

In this paper, we want to discuss the Newtonian limit of a class
of scalar--tensor theories of gravity, the induced gravity
theories, which are invoked in order to solve several problems in
cosmology (e.g. extended inflation \cite{la})  and in fundamental
physics (e.g. the tree--level--string--dilaton effective action can be recast
as an induced  gravity theory \cite{tseytlin}). In particular, we
show that the Newtonian limit of an induced gravity theory
depends on the parameters of the nonminimal coupling between
scalar field and Ricci scalar and the self--interaction potential
of such a scalar field.  Furthermore, a quadratic correction of
the Newtonian potential strictly depends on the presence of the
scalar--field potential which acts, in the low energy limit, as a
cosmological constant (Sec.2).

In Sec.3, we obtain a static, spherically symmetric,
asymptotically flat exact solution which is the generalization,
for the nonminimally coupled theory with  coupling of the form
$F(\phi)=\xi\phi^2$, of the well--known Janis--Newman--Winicour
solution to the Einstein--massless scalar equations \cite{janis}.
The deflection angle and some microlensing observables are deduced
in Sec.4. Conclusions are drawn
in Sec.5.

\section{\bf The field equations and the linearized solutions}
\setcounter{equation}{0}

The most general  action for a theory of gravity where a scalar--field
is nonminimally coupled to the geometry is, in four dimensions,
 \beq\lb{41}
 {\cal A}=\vol\left[\f R+\frac{1}{2}\, g\umunu \phi\ddemu \phi\ddenu-
 V(\phi)+{\l}_{m}\right]\,,
 \eeq
where $\f$,
$\v$ are generic functions of $\p$ and ${\l}_{m}$ is the
ordinary matter  Lagrangian density \cite{cqg},\cite{cimento}.

We obtain the field equations by varying the action with respect
to the metric tensor  field $\gd$
 \beq \lb{42}
 G\dmunu=T^{(eff)}\dmunu\,,
 \eeq
 where $G\dmunu$ is the Einstein tensor,
 \beq\lb{43}
 G\dmunu=R\dmunu-\frac{1}{2}\,\gd R\,,
 \eeq
 and $T^{(eff)}\dmunu$, whose expression is found to be
 \beqa\lb{44}
 T^{(eff)}\dmunu= \frac{1}{\f} \left\{ -\half \, \p\ddemu \p\ddenu
 + \frac{1}{4}\, \gd g\uab \p\ddea \p\ddeb\right. &-& \half \, \gd
 \v+ \no\\ -\gd \Box \f &+&\left. \f\ddemunu -4\,\pi\tilde{G}
 \,T\dmunu \right\}\,,
 \eeqa
 is the {\it effective stress-energy tensor} containing terms of nonminimal
coupling, kinetic terms, potential of the scalar field $\p$ and
the usual stress-energy tensor of matter, $T\dmunu$, calculated
from ${\l}_{m}$. $\tilde {G}$ is a dimensional, strictly positive,
constant. In our units, with $c=1$, Einstein general relativity is
obtained when the scalar field coupling $\f$ is a constant and
$\tilde{G}$ reduces to the Newton gravitational constant $G_{N}$
\cite{cimento}.

Equations for the scalar field are found by varying the action
with respect to the same field
 \beq \lb{45}
 \Box \p -R \,\fp+\vp=0\,,
 \eeq
 where $\fp=d \f/d\p$, $\vp=d\v/d\p$ and $\Box \p
\equiv g^{\mu\nu} \phi\ddemu \phi \ddenu$. This Klein-Gordon
equation can also be obtained from the contracted Bianchi identity
\cite{cimento}.

Field equations derived from the action $(\ref{41})$ can be recast
in a Brans--Dicke equivalent form by choosing
 \beq \lb{46}
 \pv=\f,\quad \omega\left(\pv\right)=\frac{\f}{2\,\fp^2},
 \quad W\left(\pv\right)=V\left(\phi(\pv)\right) \,.
 \eeq
 Action (\ref{41}) now reads
 \beq\lb{47}
 {\cal A}=\vol \left[\pv R
 +\frac{\omega(\pv)}{\pv}\, g\umunu \pv\ddemu \pv
 \ddenu-W(\pv)+{\l}_{m}\right]\,,
 \eeq
 that is nothing else but the extension of the original Brans--Dicke
proposal (where $\omega$ is a constant), with $\omega=\omega(\pv)$
plus a potential term $W(\pv)$. By varying the new action with
respect to $\gd$ and the new scalar field $\pv$, we  obtain again
the field equations. The effective stress energy tensor in
Eq.(\ref{42}) and the Klein--Gordon equation (\ref{45}), in the
same units as before, become
 \beqa\lb{48}
 T^{(eff)}\dmunu=-\frac{4\,\pi\tilde{G}}{\pv}\,T\dmunu -
 \frac{\omega(\pv)}{\pv^2} \left(\pv \ddemu \pv \ddenu - \half\,
 \gd g\uab \pv\ddea \pv \ddeb \right)&+&\no\\
 + \frac{1}{\pv}
 \left(\pv\ddemunu -\gd \Box \pv\right)&-&\frac{1}{2\,\pv}\, \gd
 W\left(\pv\right)\,,
  \eeqa
 \beq \lb{49a}
 2\,\omega\left(\pv\right) \Box\pv-
 \frac{\omega\left(\pv\right)}{\pv}\, g\uab \pv\ddea \pv \ddeb
 -\frac{d\omega\left(\pv\right)}{d\pv}\, g\uab \pv\ddea \pv \ddeb
-\pv R +\pv\, W'\left(\pv\right)=0\,.
 \eeq
 This last equation is
usually rewritten eliminating the scalar curvature term $R$ with
the help of Eq.(\ref{48}), so that one obtains
 \beq\lb{49}
 \Box\pv=\frac{1}{3-2\,\omega(\pv)} \left(-4\,\pi \tilde{G} T- 2\,
 W\left(\pv\right)+ \pv\,W'\left(\pv\right)- \frac{d
 \omega\left(\pv\right)}{d \pv}\,  g\uab \pv\ddea \pv \ddeb
 \right).
 \eeq
 The minus sign in the denominator comes from the
sign chosen in our action (\ref{47}). Our aim now is to study the
linearized equations derived from the action (\ref{41}) or, equivalently,
from (\ref{47}).

Before starting, we need a choice for the up to now arbitrary
functions $\f$ and $\v$. A rather general choice is given by
 \beq\lb{410}
 \f=\xi \p^m\,,
 \eeq
 \beq
 \lb{411} \v=\lam \p^n\,,
 \eeq
 where $\xi$ is a coupling constant, $\lambda$ gives the self--interaction
potential strength,  $m$ and $n$ are arbitrary, for the moment,
parameters. This choice is in agreement with the existence of a
Noether symmetry in the action (\ref{41}) as discussed in
\cite{cqg},\cite{cimento}. Furthermore, several scalar--tensor
physical theories (e.g. {\it induced gravity}) admit such a form
for $F(\phi)$ and $V(\phi)$.

In order to recover the Newtonian limit, we write, as usual,
the metric tensor as
 \beq\lb{412}
 \gd=\eta\dmunu+h \dmunu\,,
 \eeq
 where $\eta\dmunu$ is the Minkoskwi metric and
$h\dmunu$ is a small correction to it. In the same way, we define
the scalar field $\psi$ as a perturbation, of the same order of
the components of $h\dmunu$, of the original field $\p$, that is
 \beq\lb{413}
 \p=\varphi_0+\psi\,,
 \eeq
where $\varphi_{0}$ is a constant of order unit. It is clear that for
$\varphi_{0}=1$ and $\psi=0$ Einstein general relativity is recovered.

To write in an appropriate form the Einstein tensor $G\dmunu$, we
define the auxiliary fields
 \beq\lb{414}
 \overline{h} \dmunu\equiv h\dmunu-\half\,\eta\dmunu h\,,
 \eeq
and
 \beq\lb{415}
 \sig\da\equiv {\overline h}_{\alp\beta,\gam}
 \eta^{\beta\gam}\,.
 \eeq
Given these definitions, to the first order in $h\dmunu$, we obtain
 \beq\lb{416}
 G\dmunu=-\half \left\{\Box_{\eta} {\overline
 h}_{\mu\nu}+\eta\dmunu
 \sig_{\alp,\beta}\eta\uab-\sig_{\mu,\nu}-\sig_{\nu,\mu}
 \right\}\,,
 \eeq
 where $\Box_{\eta}\equiv \eta\umunu \phi_{,\mu}
\phi_{,\nu}$. We have not fixed the gauge yet.

We pass now to the right hand side of Eq.(\ref{42}),
namely to the effective stress energy tensor.Up to the second order
in $\psi$, the coupling function $\f$ and the potential $\v$,
by using Eqs.(\ref{410}) and (\ref{411}), become
 \beq\lb{417}
 \f\simeq\xi \left(\varphi_0^m+m\varphi_0^{m-1}\,
 \psi+\frac{m(m-1)}{2}\varphi_0^{m-2}\,\psi^2\right)\,,
 \eeq
 \beq\lb{418}
 \v\simeq\lam \left(\varphi_0^{n}+n\,\varphi_0^{n-1}\psi+
 \frac{n(n-1)}{2}\,\varphi_0^{n-2}\psi^2\right)\,.
 \eeq

To the first order, the effective stress--energy tensor becomes
 \beq\lb{420}
 \tilde{T}\dmunu=-m \varphi_0^{2m-1}\,\eta\dmunu \Box_{\eta} \psi
 +m\, \varphi_0^{2m-1}\psi_{,\mu\nu}-\,\frac{\lambda\varphi_0^{m+n}}{2\xi}\eta\dmunu
 - (4\,\pi\tilde{G})\frac{\varphi_0^{m}}{\xi} \,T\dmunu\,,
 \eeq
 and then  the field equations are
 \beq\lb{421}
 \half \left\{\Box_{\eta} {\overline
 h}_{\mu\nu}+\eta\dmunu \sig_{\alp,\beta}\eta\uab-\sig_{\mu,\nu}-
 \sig_{\nu,\mu} \right\}=m\,\varphi_0^{2m-1} \eta\dmunu \Box_{\eta} \psi -
 m\,\varphi_0^{2m-1}\psi_{,\mu\nu}+
 \eeq
 $$
 +\,\frac{\lambda\varphi_0^{m+n}}{2\xi}\eta\dmunu +
 (4\,\pi\tilde{G})\frac{\varphi_0^{m}}{\xi}\,T\dmunu\,.
 $$
 We can eliminate the  term proportional to $\psi_{,\mu\nu}$  by choosing an
appropriate gauge. In fact, by writing the auxiliary field
$\sigma_{\alpha}$, given by Eq. (\ref{415}), as
 \beq\lb{422}
 \sig\da=m \,\varphi_0^{2m-1}\psi_{,\alp}\,,
 \eeq
 field equations (\ref{421}) read
 \beq\lb{423}
 \Box_{\eta} \overline {h}\dmunu-m\,\varphi_0^{2m-1} \eta\dmunu \Box_{\eta}
 \psi\simeq \frac{\lambda\varphi_0^{m+n}}{\xi}\,\eta_{\mu\nu}+
 (8\,\pi \tilde{G}) \frac{\varphi_0^{m}}{\xi}\, T\dmunu
 \eeq
 By defining the auxiliary field with components
$\tilde {h}\dmunu$ as
 \beq\lb{424}
 \tilde {h}\dmunu \equiv
 {\overline h}\dmunu -m \varphi_0^{2m-1}\eta\dmunu \psi\,,
 \eeq
 the field equations take the simpler form
 \beq\lb{425}
 \Box_{\eta} \tilde{h}\dmunu=\frac{\lambda\varphi_0^{m+n}}{\xi}\,\eta\dmunu+
  (8\,\pi \tilde{G})\frac{\varphi_0^{m}}{\xi}\, T\dmunu
 \eeq
 The original perturbation field $h\dmunu$ can
be written in terms of the new field as (with $\tilde{h}\equiv
\eta^{\mu\nu} \tilde{h}\dmunu$)
 \beq \lb{426}
 h\dmunu=\tilde{h}\dmunu-\half\,\eta\dmunu \tilde {h}-
 m \,\varphi_0^{2m-1}\eta\dmunu \psi\,.
 \eeq
We turn now to the Klein-Gordon Eq.(\ref{45}). $\Box_{\eta}\psi$
can be written, from the linearized Klein- Gordon equation, in
terms of the matter stress energy tensor and of the potential
term. If we calculate the scalar invariant of curvature $R=\gu
R\dmunu$ from Eq.(\ref{44}), we find
 \beq\lb{427}
 \Box\p+\frac{\fp}{\f} \left( \half\, g\uab \p\ddea \p\ddeb-2\,
 \v-3\,\Box \f-4\,\pi \tilde{G} \,T\right)+ \vp=0\,,
 \eeq
 and, to the first order, it reads
 \beq\lb{428}
 \Box_{\eta}\psi+\frac{\lambda (n-2m)(n-1)\varphi_0^{n-2}}{1-3\xi m^2\varphi_0^{m-2}}\psi=
 \frac{\lambda (2m-n)\varphi_0^{n-1}}{1-3\,\xi m^2\varphi_0^{m-2}}+
  \frac{4\,\pi \tilde{G} m^2}{(1-3\,\xi m^2\varphi_0^{m-2})\varphi_0} T\,.
 \eeq
We work in the weak-field and slow motion limits, namely we assume
that the matter stress-energy tensor $T\dmunu$ is dominated by the
mass density term and we neglect time derivatives with respect to
the space derivatives, so that $\Box_{\eta}\rightarrow -\Delta$,
where $\Delta$ is the ordinary Laplacian operator in flat
spacetime. The linearized field equations (\ref{425}) and
(\ref{428}) have, for point--like distribution of matter\footnote{ To be
precise, we can define a Schwarzschild mass of the form
$$M=\int(2T^{0}_{0}-T^{\mu}_{\mu})\sqrt{-g}d^3x \,.$$},
which is $\rho(r)=M\delta(r)$, the
following solutions:\\
for $n\neq 2m$, $n\neq 1$, we get
 \beqa
 h_{00}&\simeq & \left[(4\pi
 \tilde{G})\frac{\varphi_0^{m}}{\xi}\right]\frac{M}{r}-
 \left[\frac{4\pi\lambda\varphi_0^{m+n}}{\xi}\right]r^2-
 \left[(4\pi\tilde{G})\frac{m^2\varphi_0^{2m-2}M}{1-3\,\xi
m^2\varphi_0^{m-2}}\right]\frac{e^{-pr}}{r}
 +\nonumber \\
 & & -
 \left[\frac{4\pi m\varphi_0^{2m}}{n-1}\right]\cosh (pr)\,, \label{428a} \\
 h_{il} & \simeq & \delta_{il}\left\{\left[(4\pi \tilde{G})
\frac{\varphi_0^{m}}{\xi}\right]\frac{M}{r}+
 \left[\frac{4\pi\lambda\varphi_0^{m+n}}{\xi}\right]r^2+
 \left[(4\pi\tilde{G})\frac{m^2\varphi_0^{2m-2}M}{1-3\,\xi m^2
\varphi_0^{m-2}}\right]
 \frac{e^{-pr}}{r} \right\}+\nonumber \\
   & & -\delta_{il}\left[\frac{4\pi \varphi_0^{2m}m}{n-1}\right]\cosh (pr) \,, \label{
428b} \\
 \psi & \simeq &\left[ (4\pi\tilde{G})\, \frac{mM}{1-3\,\xi
 m^2\varphi_0}\right]\frac{e^{-pr}}{r}-\left[\frac{4\pi\varphi_0
}{n-1}\right]\cosh
 (pr)\,,\label{428c}
 \eeqa
 where the parameter $p$ is given by
 \beq\label{428d}
 p^2=\frac{\lambda (n-2m)(n-1)\varphi_0^{m-2}}{1-3\,\xi m^2\varphi_0^{m-2}}\,.
 \eeq
For $n=2m$, we obtain
 \beqa \lb{431}
 h_{00} & \simeq &\left[\frac{(4\pi\tilde{G})\varphi_0^{m}(1-4\,\xi m^2\varphi_0^{m-2})}{\xi
 (1-3\,\xi m^2\varphi_0^{m-2})}\right]\frac{M}{r}-\left[\frac{4\pi
\lambda\varphi_0^{m+n}}{\xi}\right]r^2
 -\Lambda\,,\\
 \lb{432}
 h_{il} & \simeq &
 \delta_{il}\left\{\left[\frac{(4\pi\tilde{G})\varphi_0^{m}(1-2\xi m^2\varphi_0^{m-2})}{\xi
 (1-3\,\xi m^2\varphi_0^{m-2})}\right]\frac{M}{r}+
 \left[\frac{4\pi\lambda\varphi_0^{m+n}}{\xi}\right]r^2+
\Lambda \right\} \,, \\
 \lb{433}
 \psi & \simeq & \left[\frac{(4\pi\tilde{G})m}{(1-3\,\xi
m^2\varphi_0^{m-2})\varphi_0}\right]
 \frac{M}{r}+\psi_0
 \eeqa
 where $\Lambda=m\varphi_0^{2m-1}\psi_0$ and $\psi_0$ are arbitrary
integration constants. Let
us note that the values $m=1$, $n=2$ and $m=2$, $n=4$ correspond
to the well known couplings and potentials, i.e. $F\sim \phi$,
$V\sim \phi^2$ and $F\sim \phi^2$, $V\sim \phi^4$, respectively.

Finally for
 $n=1$, we obtain
 \beq \lb{433a}
 h_{00}  \simeq \left[\frac{(4\pi\tilde{G})\varphi_0^{m}(1-4\,\xi m^2\varphi_0^{m-2})}{\xi
 (1-3\,\xi m^2\varphi_0^{m-2})}\right]\frac{M}{r}-
\left[\frac{4\pi\lambda\varphi_0^{m+n}[1-\xi m(m+1)\varphi_0^{m-2}]}
 {\xi(1-3\xi m^2\varphi_0^{m-2})}\right]r^2,
\eeq
\beq
 \lb{433b}
 h_{il}  \simeq
 \delta_{il}\left\{\left[\frac{(4\pi\tilde{G})\varphi_0^{m}(1-2\,\xi m^2\varphi_0^{m-2})}{\xi
 (1-3\xi m^2\varphi_0^{m-2})}\right]\frac{M}{r}+
\left[ \frac{4\pi\lambda\varphi_0^{m+n}
 [1-\xi m(m+1)\varphi_0^{m-2}]}{\xi(1-3\xi
m^2\varphi_0^{m-2})}\right]r^2\right\},
\eeq
\beq
 \lb{433c}
 \psi  \simeq \left[\frac{(4\pi\tilde{G})m}{(1-3\xi
 m^2)\varphi_0}\right]\frac{M}{r}+
 \left[\frac{4\pi\lambda (2m-1)\varphi_0^{n-1}}{1-3\xi m^2\varphi_0^{m-2}}
\right] r^2 \,.
 \eeq
If we demand the $(0,0)$--component of the field Eq.(\ref{425}),
when $\lam=0$, to read as the usual Poisson equation (that is
nothing else but a definition of the mass)
 \beq \lb{434}
 \Delta\Phi=4\,\pi  G_{N} \rho\,,
 \eeq
 where $\Phi$, linked with the metric tensor by the relation
$h_{00}=2\,\Phi$, is the Newtonian potential, we have to put
 \beq\lb{435}
 G_{N}=-\frac{\varphi_0^{m}}{2\,\xi}\left(\frac{1-4\,\xi m^2\varphi_0^{m-2}}
 {1-3\,\xi m^2\varphi_0^{m-2}}\right) \tilde {G}\,.
 \eeq
 We may now rewrite the nonzero components of $h\dmunu$ and the
scalar perturbed field. Let us take into account, for example,
Eqs. (\ref{431})--(\ref{433}). We get
 \beq\lb{436}
 h_{00}\simeq -\frac{2\,G_{N} M}{r}- \frac{4\pi\lambda\varphi_0^{m+n}}{\xi}\,r^2
 -\varphi_0^{2m-1}\psi_0
 \eeq
 \beq\lb{437}
 h_{il}\simeq \delta_{il}\left\{-\frac{2\,G_{N}M}{r}\left(\frac{1-2\,\xi  m^2\varphi_0^{m-2}}
 {1-4\,\xi m^2\varphi_0^{m-2}}\right)+\frac{4\pi\lambda\varphi_0^{m+n}}{\xi}\,r^2
 +m\varphi_0^{2m-1}\psi_0\right\} \,,
 \eeq
 and
 \beq\lb{438}
 \psi=-\frac{2\,G_{N} M}{r} \left( \frac{\xi m\varphi_0^{-m-1}}{1-4\,\xi
 m^2}\right) +\psi_0\,,
 \eeq
where the Newton constant explicitly appears. Similar considerations hold
in the other cases.

What we have obtained are solutions of the linearized field equations, starting from the
action of a scalar field nonminimally coupled to the geometry, and minimally coupled to
the ordinary matter. Such solutions  depend on the parameters which characterize the
theory:  $\xi, m,n,\lambda$. The results of Einstein general relativity are obtained for
$\f=F_{0}$, with $F_{0}$ negatively--defined due to the sign choice in the action
(\ref{41}). As we can easily see from above, in particular from
Eqs.(\ref{436})--(\ref{438}), we have the usual Newtonian potential and a sort of {\it
cosmological term} ruled by $\lambda$ which, from the Poisson equation, gives a
quadratic contribution.

We consider now the Brans-Dicke-like action (\ref{47}) where
$\omega=\omega\left(\pv\right)$. It is actually simple to see,
from the field Eqs.(\ref{48}) and (\ref{49}), that, if we want to
limit ourselves to the linear approximation, we may consider as
well $\omega=$constant. The link with the results that follows
from the action (\ref{47}) are given by the transformation laws
(\ref{46}), that is
 \beq \lb{439}
 \omega\left(\pv(\p)\right)=\frac{1}{2\,\xi m^2} \, \p^{2- m}\,.
 \eeq
 The potential term $W\left(\pv\right)$ in the linear
approximation, behaves as $V(\phi)$. Results in the approaches given by the actions
(\ref{41}) and (\ref{47}) are completely equivalent.

\section{\bf An exact solution}
\setcounter{equation}{0}

We are going now to consider a particular choice for the
coupling field $\f$ and the potential $\v$ and to obtain an
exact solution for the field equations in the case of
the external field of a distribution of matter endowed with
spherical symmetry. Namely, we consider
 \beq\lb{31}
 \f=\xi \p^2\,,
 \eeq
 \beq\lb{32}
 \v=0\,.
 \eeq
The physical relevance of the
above assumption  is widely discussed in literature (see e.g. \cite{cimento}
and references therein). Here we want to show that it is possible
to find an exact solution of the same form of those (more general) discussed
in previous section. This is nothing else but a generalization to the
nonminimal coupling case of the well known solution of Janis--Newman--Winicour
\cite{janis},\cite{virbhadra}.

The exact solution that we will find is linked, of course, with
the one obtained in Brans-Dicke theory with $\omega=$constant
(from (\ref{439}) it follows that it is just for the value
$m$=2, that we find $\omega=$constant).

We look for a solution for a static scalar field, so that,
given  the symmetry of the problem, the Birkoff theorem
holds, and we can write the line element
as
 \beq\lb{33}
 ds^2=a(r)\, dt^2-b(r)\, dr^2-c(r)\, r^2 d\Omega^2\,,
 \eeq
 where $a(r), b(r)$ and $c(r)$ are strictly positive
functions of the radial coordinate $r$, $d\Omega^2$ is the
usual spherical element.

We can write down the Einstein field equations
(\ref{42}) as
 $$
 -\frac{a}{r^2 b}+\frac{a}{r^2
 c}+\frac{a b'}{r b^2}- \frac{3\, a c'}{r b c}+\frac{a b'
 c'}{2\,
 b^2  c}+ \frac{a c'^2}{4\,b c^2}-\frac{a c''}{b c}= \no \\
 $$
 \beq\lb{35a}
 =\frac{4\,a \p'}{r b \p}-\frac{a b' \p'}{b^2 \p}+\frac{2\,a c'
 \p'}{b c \p}+\frac{2\,a \p'^2}{b \p^2}-\frac{a \p'^2}{4\,
 \xi b
 \p^2}+\frac{2\,a \p''}{b \p}\,,
 \eeq
 \beq\lb{35b}
 \frac{1}{r^2}-\frac{b}{r^2 c}+\frac{a'}{r a}+\frac{c'}{r c}
 +\frac{a' c'}{2\,a c}+\frac{c'^2}{4 c^2}= -\frac{4\,\p'}{r
 \p}-\frac{a' \p'}{a \p}-\frac{2\,c' \p'}{c \p}-
 \frac{\p'^2}{4\,\xi \p^2}\,,
 \eeq
 $$
 \frac{r c a'}{2\,a b}-\frac{r^2 c a'^2}{4\,a^2 b}-\frac{r c
 b'}{2 b^2}-\frac{r^2 c a' b'}{4 a b^2}+\frac{r c'}{b}+\frac{r^2 a'
 c'}{4\, a b}-\frac{r^2 b' c'}{4\,b^2}- \frac{r^2 c'^2}{4\,b
 c}+\frac{r^2 c a''}{2\,a b}+\frac{r^2 c''}{2\, b}=
 $$
 \beq\lb{35c}
 =-\frac{2\,r c \p'}{b \p}-\frac{r^2 c a' \p'}{a b
 \p}+\frac{r^2 c b' \p'}{b^2 \p}-\frac{r^2 c' \p'}{b \p}-
 \frac{2\,r^2 c \p'^2}{b \p^2}+\frac{r^2 c \p'^2}{4\,\xi b
 \p^2}-\frac{2\,r^2 c \p''}{b \p}\,,
 \eeq
 where the prime now indicates the derivative with respect to $r$.
The Klein-Gordon equation is
 \beq\lb{35f}
 \left(1-12\,\xi\right)\left(\Box \p-\frac{\p'^2}{b\p}
 \right)=0\,. \eeq The case $\xi=1/12$ is the conformal
case. Of course,  the usual Schwarzschild solution of Einstein
general relativity is recovered for $\p=$constant.

We look for a solution of the form
 \beq\lb{36}
 ds^2=\left(1-\frac{\chi}{r}\right)^{\alpha} dt^2-
 \left(1-\frac{\chi}{r}\right)^{\beta} dr^2- \left(1-
 \frac{\chi}{r}\right)^{\nu} r^2 d\Omega^2\,, \eeq \beq \lb{37}
 \p\left(r\right)=\left(1-\frac{\chi}{r}\right)^{\delta}\,,
 \eeq
where $\alpha,\,\beta$ and $\nu$ are the parameters that
specify the solution (and we expect to find
$\alpha=-\beta=1$ and $\nu=0$ as a particular solution), $\chi$
is a constant related to the mass of the system which generates
the gravitational field.

Eq.(\ref{35f})  is satisfied when the algebraic relation
 \beq\lb{37a}
 4\,\delta+\alpha-\beta+2\,\nu-2=0\,,
 \eeq
 holds. By studying Einstein equations, we find that a solution of the form
(\ref{36}) does exist for
 \beq \lb{38}
 \alpha=-2\,\delta+\epsilon\,,
 \eeq
 \beq \lb{39}
 \beta=-2\,\delta-\epsilon\,,
 \eeq
 \beq \lb{38a}
 \gam=-2\,\delta-\epsilon+1\,,
 \eeq
 where $\epsilon$ is an auxiliary parameter which we discuss below.
The link between them and the coupling constant $\xi$ is the
quadratic algebraic equation
 \beq\lb{310}
 \delta^2 \left(1-12\,\xi \right)+\xi \left(1-\epsilon^2\right)=0\,.
 \eeq
 We notice that, in order to specify completely the solution, we have to
give, besides $\chi$, the value of two of the three parameters
$\delta,\,\epsilon$ and $\xi$. It is only to recover the
Schwarzschild solution that we need to specify $\epsilon^2=1$. We
have also to note that, in the weak field limit, that we are now
going to study, the parameters $\delta$ and $\epsilon$ can be
expressed in term of $\xi$.

We may study the range of applicability of the found solution
writing the Ricci scalar of curvature whose expression
is given by
 \beq \lb{311a}
 R=\left[\frac{1-\epsilon^2}{1-12\,\xi}\right]\left[\frac{ \chi^2
 \left(1-\frac{\chi}{r}\right)^{2\, \delta+\epsilon}}{2\,r^4
 \left(1-\frac{\chi}{r} \right)^2}\right]\,.
 \eeq
 We find, beside the usual
singularity in $r=0$, a singularity in $r=\chi$, which is a null
surface (see also \cite{virbhadra}). Actually $\chi$ can take both
positive and negative values, so that the solution is defined
respectively for $r> \chi$ and $r> 0$.

We now expand the solution found at the first order in $\chi/r$
and we identify it with the result of the linearized field
equations for a point-like distribution of matter, with mass $M$.
We demand the Poisson equation to hold, that is, we are taking
into account a particular case of what we considered in the last
section. We find
 \beq\lb{311}
 \chi=2\,\epsilon \,G_{N} M\,,
 \eeq
and from Eq.(\ref{435}), we have (for $\varphi_0=1,$ and $m=2$)
 \beq\lb{312}
 G_{N}=-\frac{1}{2\xi}\left(\frac{1-16\,\xi}{1-12\,\xi}\right)
 {\tilde G}\,.
 \eeq
 $\tilde {G}$, as above, is  a dimensional strictly positive
constant. Moreover we find, from the $(0,0)$--component of the
field Eq. (\ref{425}) and Eq.(\ref{428}) for the scalar
perturbation field $\psi$, where we put $m=2$ and $\lam=0$, a
second equation for $\delta,\,\epsilon$ and $\xi$,
 \beq \lb{313}
 \left(1-16\,\xi\right) \delta=2 \left(\epsilon-2\,\delta \right)
 \xi\,.
 \eeq
 Together with Eq.(\ref{311}), this equation allows us
to express the solution as a function of the parameter $\xi$
alone. It results
 \beq\lb{314}
  \epsilon=
 \sqrt{\frac{1-12\,\xi}{1-16\,\xi}}\,,
 \eeq
 and
 \beq\lb{315}
 \delta=\frac{2\,\epsilon\,\xi}
 {1-12\,\xi}=\frac{2\, \xi} {\sqrt{(1-12\,\xi)(1-16\,\xi)}}\,.
 \eeq
 As it can be seen, the allowed range for $\xi$ is $\xi <1/16$,
$\xi > 1/12$ (from which one should exclude the value $\xi=0$, the
value that would give $\delta=0$ and $\epsilon^2=1$). The role of
the parameter $\epsilon$ is now clarified: it tells us how much
the solution differs from the usual Schwarzschild solution of
general relativity since we are considering the nonminimal coupling
$F(\phi)=\xi\phi^2$.

The linearized solution can then be written as
 \beqa\lb{316}
 ds^2&=&\left(1-\frac{2\,{ G_{N}} M}{r}\right) dt^2-\left[1-
 \frac{2\,{G_{N}}
 M}{r}\,\left(\frac{1- 8\,\xi} {1-16\,\xi}\right) \right] dr^2+\no\\
 &-&\left[1-\frac{2\,{G_{N}} M}{r}\left(\frac{1-8\,\xi} {1-16\,\xi}-
 \sqrt{\frac{1-12\,\xi}{1-16\,\xi}}\right)\right]\,  r^2
 d\Omega^2,
 \eeqa
 \beq\lb{317}
 \p\left(r\right)=1-\frac{2\,{G_{N}} M}{r} \,
 \left(\frac{2\,\xi}{1-16\,\xi} \right)\,,
 \eeq
which, by a linear
transformation for the radial coordinate, can be put in the same
form of the solution (\ref{436}), (\ref{437}) and (\ref{438}) with
the choice $m=2$ and $\lam=0$.

In order to recover the corresponding Brans-Dicke solution,
with $\omega=$constant and $\p\rightarrow \pv=\xi \p^2$,
let us introduce the coordinate transformation
 \beq \lb{318}
 \left(\frac{dr}{dr'}\right)^2=\left(\frac{r}{r'}\right)^2
 \frac{c(r)}{b(r)}= \left(\frac{r}{r'}\right)^2
 \left(1-\frac{\chi}{r}\right)^{\nu-\beta}=
 \left(\frac{r}{r'}\right)^2 \left(1-\frac{\chi}{r}\right)\,,
 \eeq
 where, as already noted, $\beta=-1$ and $\nu=0$.
The new variable
 \beq \lb{318a}
 r=r'\left(1+\frac{\chi}{4\,r'}\right)^2\,,
 \eeq
 is consistent with (\ref{318}). The line element
in the new coordinate system is
 \beq \lb{318b}
 ds^2=\left(\frac{1-\chi/r'}{1+\chi/r'}\right)^{2\,\alpha}
 dt^2-\left(1+\frac{\chi}{r'}\right)^4 \left(\frac{1- \chi/r'}
 {1+\chi/r'}\right)^{2\,\beta+2} \left(dr'^2+r'^2
 d\Omega^2\right)\,,
 \eeq
 while the scalar field is
 \beq \lb{318c}
 \p\left(r'\right)= \left(\frac{1-\chi/r'}
 {1+\chi/r'}\right)^{2\,\delta}\,.
 \eeq
 The Brans-Dicke solution is then  immediately found
(see also \cite{romero} for the asymptotic discussion). The line
element is the same as (\ref{318b}), whereas the scalar field is
equal to
 \beq\lb{318d}
 \pv\left(r'\right)= \frac{1}{8\,\omega}
 \left(\frac{1- \chi/r'} {1+\chi/r'}\right)^{4\,\delta}\,.
 \eeq

If we introduce the new parameters
$\epsilon'$, $\delta'$ and $\nu'$
 \beq\lb{323}
 2\,\alpha\equiv\alpha'\equiv\frac{2}{\epsilon'},\quad 2\,
 \beta+2\equiv\beta'\equiv\frac{2} {\epsilon'} \,
 \left(\epsilon'-\delta'-1\right),\quad
 2\,\nu\equiv\nu'\equiv 4\,\delta\,,
 \eeq
 we find
 \beq\lb{324}
 \nu'=\frac{\delta'}{\epsilon'}\,.
 \eeq
 As it has been done for  $\xi$, it is possible to express everything
as a function of the parameter $\omega$.

It is worth noticing that the above solutions are equivalent, up
to a conformal transformation, to the Schwarzschild-like
solution obtained in the contest of  minimally coupled
theory of gravitation with a scalar field given by the action
 \beq\lb{328}
 {\cal A}=\int d^{4}x\sqrt{-\tilde {g}}\left[k \tilde{R}
 +\frac{1}{2} \tilde{g}\umunu \tilde{\phi}\ddemu
 \tilde{\phi}\ddenu\right]\,,
 \eeq
 where $k$ is a dimensional, strictly negative, constant that
fixes units and $\sqrt{-\tilde{g}}$ is the square root of the determinant of conformally
transformed metric. This solution is known as the Janis--Newman--Winicour one of the
Einstein--massless scalar field theory \cite{janis}.

However, we have to stress that, when we look for the
linearized equations, such a conformal transformation has to be
performed, if we want to express the parameters that specify the
solution as a function of the physical quantities of the system.
For example, to define the mass, we must choose the coupling with
the ordinary matter. That is, we must decide whether the {\it
Jordan frame} description assumed for the action (\ref{41}) or
(\ref{47}) is the physical one, or if, by coupling minimally to
the scalar field also the Lagrangian for ordinary matter in action
(\ref{328}), the {\it Einstein frame} description is the physical
one. Of  course, in the two case  we get different results. For a
detailed discussion, see e.g. \cite{cqg1}. Furthermore, deviations
from the usual expression for the bending angle of light in
Einstein general relativity, are already present at first order
for solutions in the Jordan frame, but only at second order in the
Einstein frame \cite{pireaux}. In this sense the conformal
equivalence of  Schwarzschild-like solutions in the two frames is
broken when we introduce ordinary matter.

\section {\bf Deflection Angle and Microlensing Observables}
\setcounter{equation}{0}

In the previous sections, we dealt with the problem of
the determination of the geometry of a static and
spherically symmetric spacetime in a given nonminimally coupled
theory of gravity.
From this solution it is possible to
calculate, in a straightforward way,  the deflection of light. We
start with a line element written in the form
 \beq \lb{21}
 ds^2=a\left(r\right) dt^2-b\left(r\right) \left(dr^2+r^2
 d\Omega^2\right)\,,
 \eeq from which the geodesic equation of
motion for photons is found to be
 \beq \lb{22}
 u^2\left(u_{\hat{\phi}\hat{\phi}}+u\right)+\half \left(b^{-1} b'-a^{-1}
 a'\right)\left(u^2+u^2_{\hat{\phi}}\right)=0\,,
 \eeq
 where $u\equiv 1/r$ and $u_{\hat{\phi}}\equiv du/d\hat{\phi}$
(here $\hat{\phi}$ is the azimuthal angle). If we consider this
equation in the same limit where our solution (\ref{436}),
(\ref{437}) is valid, we get (putting $\psi_0=0$)
 \beq\lb{23}
 a\left(r\right)=1-2\,\frac{G_{N}M}{r}-\frac{4\pi\lambda\varphi_0^{m+n}}{\xi} r^2\,,
 \eeq
 \beq\lb{23a}
 b\left(r\right)=1+2\,\frac{G_{N}M}{r} \gamma+\frac{4\pi\lambda\varphi_0^{m+n}}{\xi} r^2\,,
 \eeq
 where $\lambda$ and $\gamma=\gamma(\xi, m)$, which is given by
 \beq\lb{24}
 \gamma=\frac{1-2\,\xi m^2\varphi_0^{m-2}}{1-4\,\xi m^2\varphi_0^{m-2}}\,,
 \eeq
 parametrize the scalar--tensor theory. We can recast Eq.(\ref{22}) as
 \beq
 u_{\phi\phi}+u=G_{N}M\left(1+\gamma\right)
 \left(u^2+u^2_\phi\right)\,.
 \eeq
 It is worth noticing
that the potential term $\lambda$ does not appear in this equation,
so that the deflection angle, $\hat\alpha$, does not depend from it
either (at least in the approximation order which we have considered).
We have
 \beq \lb{25}
 \hat\alpha=2\,\frac{G_{N}M}{r_{0} c^2} \left(1+\gamma\right),
 \eeq
 where, as usual, $r_{0}$ is the minimum distance from the deflector, and,
to this order, it is nothing else but an impact parameter. We
have reintroduced the dimensional constant $c$. The result is
similar to that given in \cite{sirousse} where also the singular
isothermal sphere model has been studied. These arguments specify how
deflection of light depends on the parameters $\xi$ and $m$  of our
scalar--tensor theory. This give us the chance of looking at the
well known effect of gravitational lensing, as a possible test for
these theories. We are now going to give the expression for some
quantities relevant for this phenomenon.

We consider the simplest situation, that is, the so called
{\it Schwarzschild lens}. With this geometry usually two images are
formed, but when they are no longer separables, and the source
brightness is magnified, one speaks about {\it microlensing}. In
this context the relevant quantity is the {\it Einstein radius},
$R_E$, that is given by
 \beq\lb{26}
 R_E=\sqrt{\frac{2\,G_{N}M}{c^2}\,\left(1+\gamma\right)\, \frac{D_l
 D_{ls}}{D_s}}\,,
 \eeq
 where $D_l$, $D_{ls}$ and $D_s$ stand
respectively for the distances between observer and lens, lens and
source, observer and source. The {\it Einstein angle}, $\theta_E$ is
given by $R_E D_l$.

The maximum amplification is equal to
 \beq\lb{27}
 A_{max}=\frac{r_{0}^2+2\,R^2_E}{r_{0} \sqrt{r_{0}^2+4\,R^2_E}}\,.
 \eeq
 Clearly, also $A_{max}$ depends on the parameters of the given scalar--tensor
theory  and, in general, all the quantities of gravitational lensing
can be {\it parametrized} by them. In conclusion, beside classical tests
\cite{will}, gravitational lensing could be useful, in general, to test
gravity. On the other hand, anomalies in lensing features could be explained
by enlarging the available set of gravitational theories.

It is interesting to observe that Eq. (\ref{24}) allows to
determine the relation between $\xi$ and $m$. In fact, inverting
(\ref{24}) one gets
 \beq\lb{28}
 \xi m^2= \frac{1-\gamma}{2(1-2\gamma)}\,,
 \eeq
considering, for simplicity, $\varphi_{0}=1$.  A measurement of $\gamma$,
i.e. the deviation from the expected value of the gravitational mass
$G_{N}M$, estimated for example by Keplerian motions, could roughly indicate
the form of the relativistic theory of gravity which holds at a given scale.
On the other hand, the result strictly depends on the accuracy by which
$\hat{\alpha}$ is measured.

\section{\bf Discussion and Conclusions}

The Newtonian limit of induced--gravity theories, where the scalar
field is nonminimally coupled to the geometry, strictly depends on
the parameters
of the coupling and the self--interaction potential as it has to be
by a straightforward PPN parametrization \cite{will}.

In this paper, we have constructed the weak limit solution of
induced--gravity assuming, as it is quite natural to do, power law couplings
and potentials.

As we have seen, the role of the self--interaction potential is essential to obtain
corrections to the Newtonian  potential. Such corrections are, in any case, constant,
quadratic or Yukawa--like as for other generalized theories of gravity
\cite{stelle,mannheim,kenmoku}, so also induced gravity could be a candidate to solve
the problems of flat rotation curves of spiral galaxies [see e.g. the solution
(\ref{428a})]. Essentially the corrections depends on the strength of the coupling and
the ``mass" of the scalar field $\phi$ given by $\lambda$. Besides, we have always scale
lengths where Einstein general relativity can be recovered. This fact could account why
measurements {\it inside} the Solar System confirm such theory  while {\it outside} of
it there are probable deviations \cite{anderson}. Furthermore, it is possible to show
that the {\it induced-gravity picture} and the {\it Brans-Dicke picture} are completely
equivalent in the Newtonian limit as it trivially has to be.

The so called Janis--Newman--Winicour solution of minimally coupled
scalar--tensor gravity can be easily extended to nonminimally coupled
theories (at least without self--interaction potential) and it appears as
a particular case of the general Newtonian solution (\ref{436})--(\ref{438})
discussed above.

Finally all the lensing observables, first of all the deflection angle, are
affected if we introduce a nonminimal coupling and, consequently,
gravitational lensing  could constitute a further tool to investigate
relativistic theories of gravity. Vice versa,  relativistic theories could
explain anomalous effects in gravitational lensing.

In the case which we have analysed, it is interesting to stress the fact that the
self--interaction potential of the scalar field does not intervene while deviations and
then the  ``parametrization" strictly depend on the coupling.


\begin{thebibliography}{99}
\bibitem{starobinsky} A.A. Starobinsky, \pl {\bf 91B} (1980) 99.
\bibitem{la} D. La and P.J. Steinhardt, \prl {\bf 62} (1989) 376.
\bibitem{weinberg} S. Weinberg, {\it Gravitation and Cosmology}, Wiley, 1972
                   New York N.Y.
\bibitem{vilenkin} A. Vilenkin, \pr {\bf 32 D} (1985) 2511.
\bibitem{odintsov} I.L. Buchbinder, S.D. Odintsov, and I.L. Shapiro,
                   {\it Effective Action in Quantum Gravity}, IOP Publishing
                   (1992) Bristol.
\bibitem{will} C.M. Will, {\it Theory and Experiments in Gravitational
               Physics} (1993) Cambridge Univ. Press, Cambridge.
\bibitem{stelle} K. Stelle, \grg {\bf 9} (1978) 353.
\bibitem{sanders} R.H. Sanders, \araa {\bf 2} (1990) 1.
\bibitem{mannheim} P.D. Mannheim and D. Kazanas, \apj {\bf 342} (1989) 635.
                   O.V. Barabash and Yu. V. Shtanov, \pr {\bf 60 D}
                   (1999) 064008.
\bibitem{kenmoku} M. Kenmoku, Y. Okamoto, and K. Shigemoto, \pr {\bf 48 D}
                  (1993) 578.
\bibitem{anderson} J.D. Anderson et al. \prl {\bf 81} (1998) 2858.
\bibitem{schmidt} I. Quant and H.-J. Schmidt, {\it Astron. Nachr.} {\bf 312}
                  (1991) 97.
\bibitem{ehlers} P. Schneider, J. Ehlers, and E.E. Falco, {\it Gravitational
                Lenses} Springer--Verlag (1992) Berlin.
\bibitem{krauss} L.M. Krauss and M. White, \apj {\bf 397} (1992) 357.
\bibitem{nottale} L. Nottale in {\it Dark Matter (Moriond Astrophysics
                  Meetings)}, J. Andouze and J. Tran Thanh Van eds.
                  (1988) Frontieres, Gif--sur--Yvette.
\bibitem{tseytlin} A.A. Tseytlin and C. Vafa, \np {\bf 372B} (1992) 443.
\bibitem{janis} A.I. Janis, E.T. Newman, J. Winicour, \prl {\bf 20} (1968)
                878.
\bibitem{cqg} S. Capozziello and R. de Ritis, \pl {\bf 177 A} (1993) 1.\\
              S. Capozziello and R. de Ritis, \cqg {\bf 11} (1994) 107.
\bibitem{cimento} S. Capozziello, R. de Ritis, C. Rubano, and P. Scudellaro
                  {\it La Rivista del Nuovo Cimento} {\bf 4} (1996) 1.
\bibitem{virbhadra} K.S. Virbhadra, Narasimha, and S.M. Chitre, \aa {\bf 337}
                    (1998) 1.
\bibitem{romero} C. Romero and A. Barros, \pl {\bf 173A} (1993) 243.\\
                 N. Banerjee and S. Sen, \pr {\bf 56D} (1997) 1334.
\bibitem{cqg1} G. Magnano and L. M. Sokolowski, \pr {\bf 50D} (1994) 5039.\\
              S. Capozziello, R. de Ritis, and A.A. Marino, \cqg {\bf 14}
              (1997) 3243.
\bibitem{pireaux} J.-M. Gerard and S. Pireaux, gr-qc/9907034 (1999).
\bibitem{sirousse} H. Sirousse--Zia, \grg {\bf 30} (1998) 1273.



\end{thebibliography}
\end{document}